\documentclass[aps,prb,preprint,superscriptaddress]{revtex4-1}

\usepackage{graphicx}
\usepackage{dcolumn}
\usepackage{bm}
\usepackage[breaklinks]{hyperref}
\usepackage[all]{hypcap}
\usepackage{color}
\usepackage{array}

\newcommand{\ff}{\frac{1}{2}}

\begin{document}


\title{Quantum oscillations of a linear chain of coupled orbits with small effective masses: the organic metal $\theta$-(BETS)$_4$CoBr$_4$(C$_6$H$_4$Cl$_2$) }



\author{Alain Audouard}
\affiliation{Laboratoire National des Champs Magn\'{e}tiques
Intenses (UPR 3228 CNRS, INSA, UGA, UPS) 143 avenue de Rangueil,
F-31400 Toulouse, France.}

\author{Jean-Yves~Fortin}
\affiliation{Institut Jean Lamour, D\'epartement de Physique de la
Mati\`ere et des Mat\'eriaux,
CNRS-UMR 7198, Vandoeuvre-les-Nancy, F-54506, France.
}%

\author{David Vignolles}
\affiliation{Laboratoire National des Champs Magn\'{e}tiques
Intenses (UPR 3228 CNRS, INSA, UJF, UPS) 143 avenue de Rangueil,
F-31400 Toulouse, France.}

\author{Rustem~B.~Lyubovskii}
\affiliation{Institute of Problems of Chemical Physics, Russian Academy of Sciences, 142432 Chernogolovka, MD, Russia}%

\author{Lo\"{\i}c Drigo}
\affiliation{Laboratoire National des Champs Magn\'{e}tiques
Intenses (UPR 3228 CNRS, INSA, UJF, UPS) 143 avenue de Rangueil,
F-31400 Toulouse, France.}

\author{Elena I. Zhilyaeva }
\affiliation{Institute of Problems of Chemical Physics, Russian Academy of Sciences, 142432 Chernogolovka, MD, Russia}

\author{Rimma N. Lyubovskaya}
\affiliation{Institute of Problems of Chemical Physics, Russian Academy of Sciences, 142432 Chernogolovka, MD, Russia}

\date{\today}


%

\begin{abstract}

De Haas-van Alphen (dHvA) and Shubnikov-de Haas (SdH) oscillations of the organic metal $\theta$-(BETS)$_4$CoBr$_4$(C$_6$H$_4$Cl$_2$) are studied in magnetic fields of up to 55 T at liquid helium temperatures. In line with Fermi surfaces (FS) illustrating the linear chain of coupled orbits, the observed Fourier components are linear combinations of the frequencies linked to the two basic orbits $\alpha$ and $\beta$, which have small effective masses compared to other organic metals with the same FS topology. Analytical formulas based on a second order development of the free energy within the canonical ensemble, not only account for the field and temperature dependence of the dHvA amplitudes but also for their relative values. In addition, strongly non-Lifshitz-Kosevich behaviours are quantitatively interpreted. In contrast, Shubnikov-de Haas oscillations are not accounted for by this model.

\textbf{short title:} Quantum oscillations of $\theta$-(BETS)$_4$CoBr$_4$(C$_6$H$_4$Cl$_2$)
\end{abstract}


\maketitle

\section{Introduction}

Many charge transfer salts based on either the bis-ethylenedithio-tetrathiafulvalene (ET) or the  bis-ethylenedithio-tetraselenafulvalene (BETS) molecule are organic metals. In many cases, their Fermi surface (FS) is an illustration of the linear chain of orbits coupled by magnetic breakdown (MB) which is the model FS proposed by Pippard to calculate MB amplitudes \cite{Pi62} (see the insert of Fig.~\ref{Fig:d_TF_FS}). The first and most famous experimental realization of this FS was provided by the organic superconductor $\kappa$-(ET)$_2$Cu(SCN)$_2$ \cite{Ur88,Os88}. In high enough magnetic fields, such FS give rise to quantum oscillations with a spectrum composed of linear combinations of the frequencies $F_{\alpha}$ and $F_{\beta}$
linked, respectively, to the closed orbit $\alpha$ and to the MB orbit $\beta$, the area of which is equal to that of the first Brillouin zone (FBZ) (for a review, see  e.g. Ref.~\onlinecite{Uj08}). The point is that, in addition to the frequencies predicted by the semiclassical model of coupled orbits network by Falicov and Stachowiak \cite{Fa66,Sh84}, 'forbidden frequencies', such as  $F_{\beta-\alpha}$ are observed in de Haas-van Alphen (dHvA) oscillations spectra. At variance with  magnetoresistance, which in addition to Shubnikov-de Haas (SdH) oscillations can evidence quantum interference (QI) linked to $e.g.$ the $\beta-\alpha$ QI path, dHvA oscillations are only sensitive to the density of states. Therefore, the $\beta-\alpha$ component should not be observed in dHvA spectra. Besides, field dependent amplitudes of few components linked to harmonics such as $2\alpha$ and MB orbits such as $\beta+\alpha$ are not in agreement with the Falicov-Stachowiak model.

In addition to $\kappa$-(ET)$_2$Cu(SCN)$_2$ \cite{Au16}, these issues have been recently addressed for $\theta$-(ET)$_4$CoBr$_4$(C$_6$H$_4$Cl$_2$) \cite{Au12} and $\theta$-(ET)$_4$ZnBr$_4$(C$_6$H$_4$Cl$_2$) \cite{Au15}. In the following, these two latter compounds are referred to as ET$_4$-Co and ET$_4$-Zn, respectively.  In short, the field and temperature dependence of the observed Fourier components are accounted for by a second order development of the free energy within the canonical ensemble, in contrast to the LK formula which only involves a first order development. As a result, Fourier amplitudes can be expressed by second order polynomials in damping factors as reported in the appendix. As an example, the amplitude of the $\beta-\alpha$ component, which do not involve any classical orbit, is accounted for by second order terms only.

Here, we consider the charge transfer salt $\theta$-(BETS)$_4$CoBr$_4$(C$_6$H$_4$Cl$_2$). SdH oscillations of this strongly two-dimensional organic metal have been studied in magnetic fields of up to 14 T \cite{Ly13}. Reported oscillatory spectra evidence frequency combinations in agreement with the above mentioned framework. The main feature of this organic metal is the small effective masses linked to the $\alpha$ and $\beta$ orbits ($m_{\alpha}$ = 1.1, $m_{\beta}$ = 1.9) which are by a factor of about three smaller than for $\kappa$-(ET)$_2$Cu(SCN)$_2$ ($m_{\alpha}$ = 3, $m_{\beta}$ = 6, see Ref.~\onlinecite{Au16} and references therein) allowing to check the model at high magnetic field with a set of parameters (effective masses and, as reported hereafter, MB field) strongly different from those of the compounds considered in previous studies. As reported in the following, unusual features are observed and nevertheless accounted for by the model.

\section{Experimental}

Crystals were synthesized by the standard electrocrystallization technique as reported in Ref.~\onlinecite{Sh11}. They were studied in pulsed magnetic fields of up to 55 T with a pulse decay duration of 0.32 s. dHvA oscillations were measured through magnetic torque measurements of a crystal with approximate dimensions 0.1 $\times$ 0.1 $\times$ 0.04~mm$^3$, stuck on a microcantilever. Variations of the microcantilever piezoresistance were measured at liquid helium temperatures with a Wheatstone bridge with an $ac$ excitation at a frequency of 63 kHz. Magnetic torque amplitudes $A^{\tau}_{\eta}$ relevant to a given Fourier component $\eta$ are related to the dHvA amplitude $A_{\eta}$ by $A^{\tau}_{\eta}$ =
$\tau_0BA_{\eta}$ where $B$ is the magnetic field and $\tau_0$ is a prefactor depending on the crystal
mass, cantilever stiffness and tilt angle $\theta$ between the field direction and the normal to the conducting plane.
Shubnikov-de Haas (SdH) oscillations were measured through contactless tunnel diode oscillator (TDO)-based method \cite{Dr10,Au12} on another crystal with approximate dimensions 1 $\times$ 1 $\times$ 0.04~mm$^3$. The angle between the normal to the conducting plane and the magnetic field direction was $\theta$ = 10$^{\circ}$ for both crystals.

\section{Results and discussion}

\begin{figure}
\centering
\includegraphics[width=0.7\columnwidth,clip,angle=0]{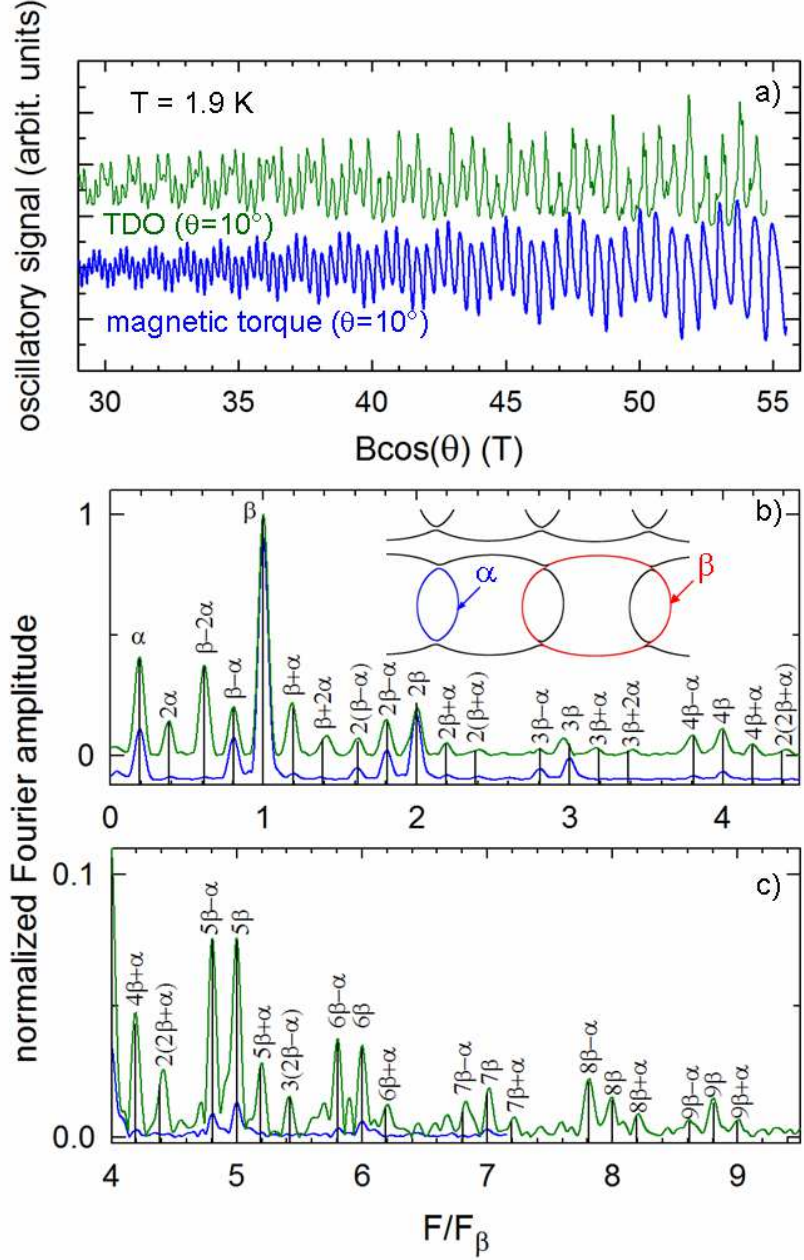}
\caption{\label{Fig:d_TF_FS} (color on line)  (a) Oscillatory part of the  TDO and torque   signal at 1.9 K and (b), (c) corresponding Fourier analysis for the field range 40-55 T (Fourier spectra are shifted down from each other for clarity). The angle between the field direction and the magnetic field is $\theta$ = 10$^{\circ}$. Thin lines in (b) and (c) are marks calculated with $F_{\alpha} (\theta=0)$ = 0.86 kT and  $F_{\alpha}/F_{\beta}$ = 0.195. The inset displays a sketch of the Fermi surface in which the basic orbits $\alpha$ and $\beta$ are marked by blue and red lines, respectively. }
\end{figure}

\begin{figure}
\centering
\includegraphics[width=0.7\columnwidth,clip,angle=0]{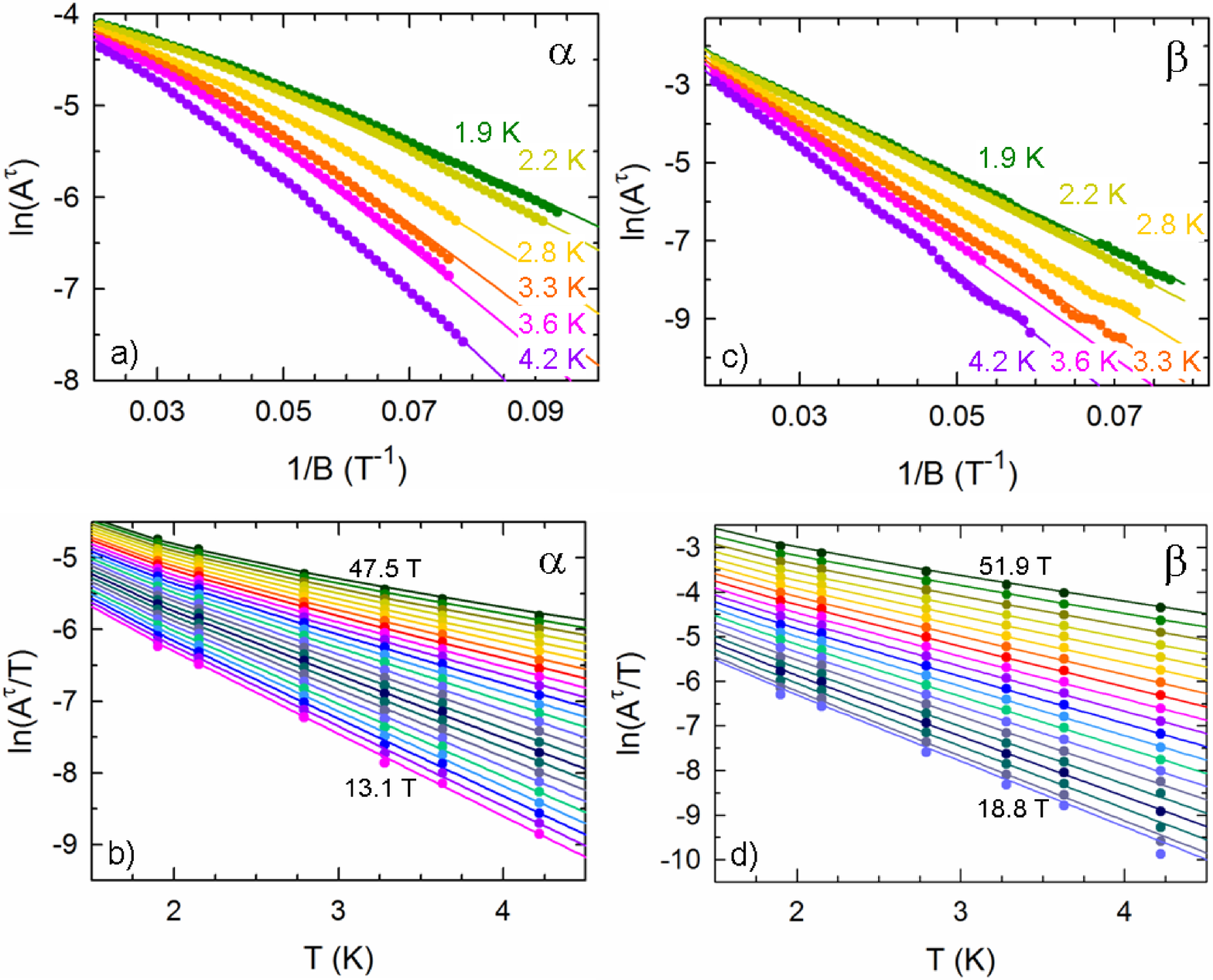}
\caption{\label{Fig:alpha_beta} (color on line)  Dingle and mass plots of (a), (b) $\alpha$ and (c), (d) $\beta$ components. Solid lines are best fits of Eqs.~\ref{Eq:alpha} and~\ref{Eq:beta}, respectively, to the data. They are obtained with $m_{\alpha}(\theta=0)$ = 1.00, $m_{\beta}(\theta=0)$ = 1.88, $B_0(\theta=0)$ = 11.6 T and $T_D$ = 0.66 K. Uncertainty on these parameters is given in the text. Data of mass plots are obtained at magnetic field values evenly spaced in 1/B within the range indicated in (a) and (c). }
\end{figure}

Field-dependent TDO and magnetic torque data at 1.9 K, along with corresponding Fourier analysis, are reported in Fig.~\ref{Fig:d_TF_FS}. Fourier spectra are composed of linear combinations of the two frequencies $F_{\alpha}$ and $F_{\beta}$, as it is the case of ET$_4$-Co \cite{Au12} and ET$_4$-Zn \cite{Au15}, the Fermi surface of which illustrate the linear chain of coupled orbits (see the insert of Fig.~\ref{Fig:d_TF_FS}). Fourier analysis yield $F_{\alpha}(\theta=0)$ = 0.860$\pm$0.004 kT and $F_{\beta}(\theta=0)$ = 4.408$\pm$0.004 kT, in agreement with low field data of Ref.~\onlinecite{Ly13}, leading to $F_{\alpha}/F_{\beta}$ = 0.195. This value is similar to those of ET$_4$-Co and ET$_4$-Zn for which $F_{\alpha}/F_{\beta}$ = 0.206 and 0.205, respectively.
Compared to data relevant to these latter $\theta$-phase compounds, an unprecedentedly large number of Fourier components can be observed, up to 6$\beta$ ($F_{6\beta}$ = 26.4 kT) and 9$\beta+\alpha$ ($F_{9\beta+\alpha}$ = 40.6 kT) for magnetic torque and TDO data, respectively.

Let us consider the magnetic torque data for which we will follow the process already adopted in Refs.~\onlinecite{Au12,Au15,Au16}. Recall that the amplitude ($A_{\eta}$) of the Fourier component with frequency $F_{\eta}=n_{\alpha}F_{\alpha}+n_{\beta}F_{\beta}$
is accounted for by analytic formulas given in the appendix. Briefly, provided the spin damping factors $R^{s}_{\alpha,1}$ and $R^{s}_{\beta,1}$ relevant to the basic components $\alpha$ and $\beta$ are not close to zero, contributions of the second order terms of Eqs.~\ref{Eq:alpha} and~\ref{Eq:beta} are negligible. As a result, these amplitudes are accounted for by the first order term, i.e. by the Lifshitz-Kosevich (LK) formula. In such a case, the spin damping factors act as temperature- and field-independent prefactors. Nevertheless, five independent parameters still enter the amplitudes: the effective masses $m_{\alpha}$ and $m_{\beta}$, Dingle temperatures $T_{D\alpha}$ and $T_{D\beta}$ and the MB field $B_0$. For this reason, it is further assumed that the Dingle temperature is the same for both orbits ($T_{D\alpha} = T_{D\beta} = T_D$). These parameters having been determined from the data relevant to $\alpha$ and $\beta$, the effective Land\'{e} factors $g_{\alpha}$ and $g_{\beta}$ can be determined from the data relevant to frequency combinations \cite{Au16} or angle dependence of the amplitudes \cite{Au15}.

Field and temperature dependence of the $\alpha$ and $\beta$ components amplitude is reported in Fig.~\ref{Fig:alpha_beta}. Best fits to the data yield $m_{\alpha}(\theta=0)$ = 1.00$\pm$0.05, $m_{\beta}(\theta=0)$ = 1.88$\pm$0.08 (in $m_e$ units), $B_0(\theta=0)$ = 11.6$\pm$3.2 T and $T_D$ = 0.66$\pm$0.10 K. In agreement with the low field data of Ref.~\onlinecite{Ly13}, effective mass values are very small compared to other organic metals with the same FS topology. MB field is significantly lower than for  ET$_4$-Co ($B_0$ = 35$\pm$5 T)  and ET$_4$-Zn  ($B_0$ = 26$\pm$3 T), as well. Combination of small effective masses and MB field is certainly responsible for the very large number or frequency combinations observed in the data of  Fig.~\ref{Fig:d_TF_FS}.

\begin{figure}
\centering
\includegraphics[width=0.9\columnwidth,clip,angle=0]{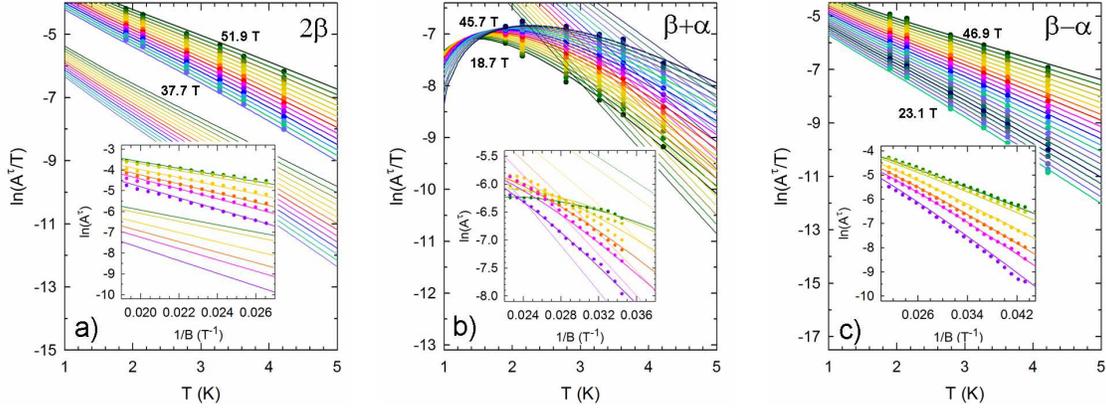}
\caption{\label{Fig:combinaisons} (color on line) Mass plots of (a)  2$\beta$, (b) $\beta+\alpha$, and (c) $\beta-\alpha$ Fourier components. Data are evenly spaced in $1/B$ within the indicated field values. The inserts display the corresponding Dingle plots. Solid lines are best fits of Eqs.~\ref{Eq:2beta},~\ref{Eq:beta+alpha} and \ref{Eq:beta-alpha}, respectively, to the data obtained with $m_{\alpha}(\theta=0)$ = 1.00, $m_{\beta}(\theta=0)$ = 1.88, $B_0(\theta=0)$ = 11.6 T and $T_D$ = 0.66 K (which are the same values as those deduced from the $\alpha$ and $\beta$ components, see Fig.~\ref{Fig:alpha_beta}) and $g_{\alpha}$ = $g_{\beta}$ = 1.85. Thin lines correspond to the contribution of the first order term (i.e. the Lifshitz-Kosevich formula). }
\end{figure}

\begin{figure}
\centering
\includegraphics[width=0.7\columnwidth,clip,angle=0]{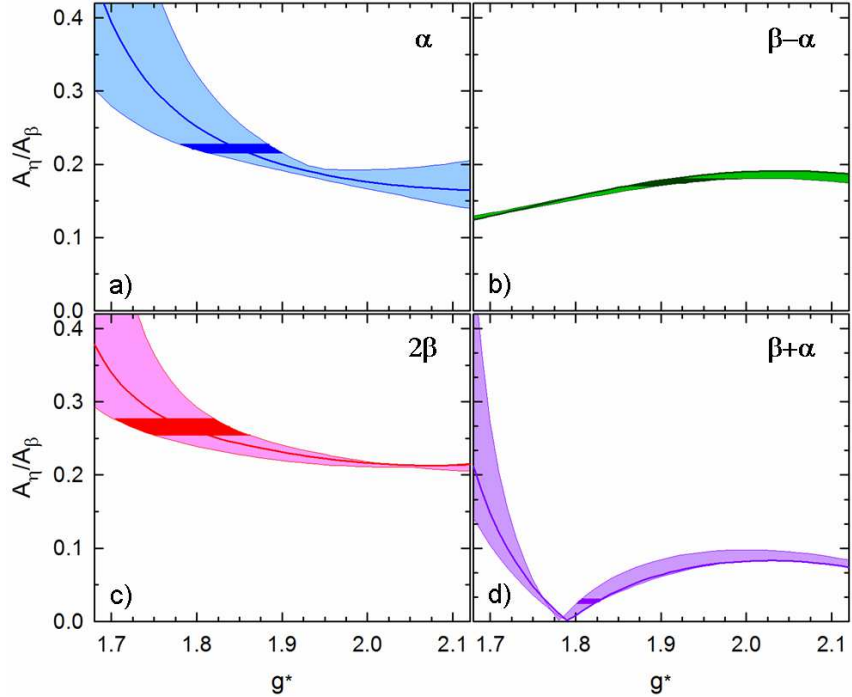}
\caption{\label{Fig:Aeta_Abeta} (color on line)  Influence of the effective Land\'e factor ($g^*=g_{\alpha}=g_{\beta}$, see text) on Fourier amplitudes $A_{\eta}$ normalized to $A_{\beta}$ for T=1.9 K and B=45 T. Solid lines are deduced from Eqs. (a)~\ref{Eq:alpha}, (b)~\ref{Eq:beta-alpha}, (c)~\ref{Eq:2beta} and (d)~\ref{Eq:beta+alpha} with the same parameters as in Figs.~\ref{Fig:alpha_beta} and~\ref{Fig:combinaisons} ($A_{\beta}$ is given by Eq.~\ref{Eq:beta}). Lightly shaded areas accounts for the uncertainty on these parameters given in the text. Heavily shaded areas stand for experimental data, taking into account the experimental uncertainty: effective  Land\'e factors in the range 1.7-2.0 account for these data. }
\end{figure}

\begin{figure}
\centering
\includegraphics[width=0.7\columnwidth,clip,angle=0]{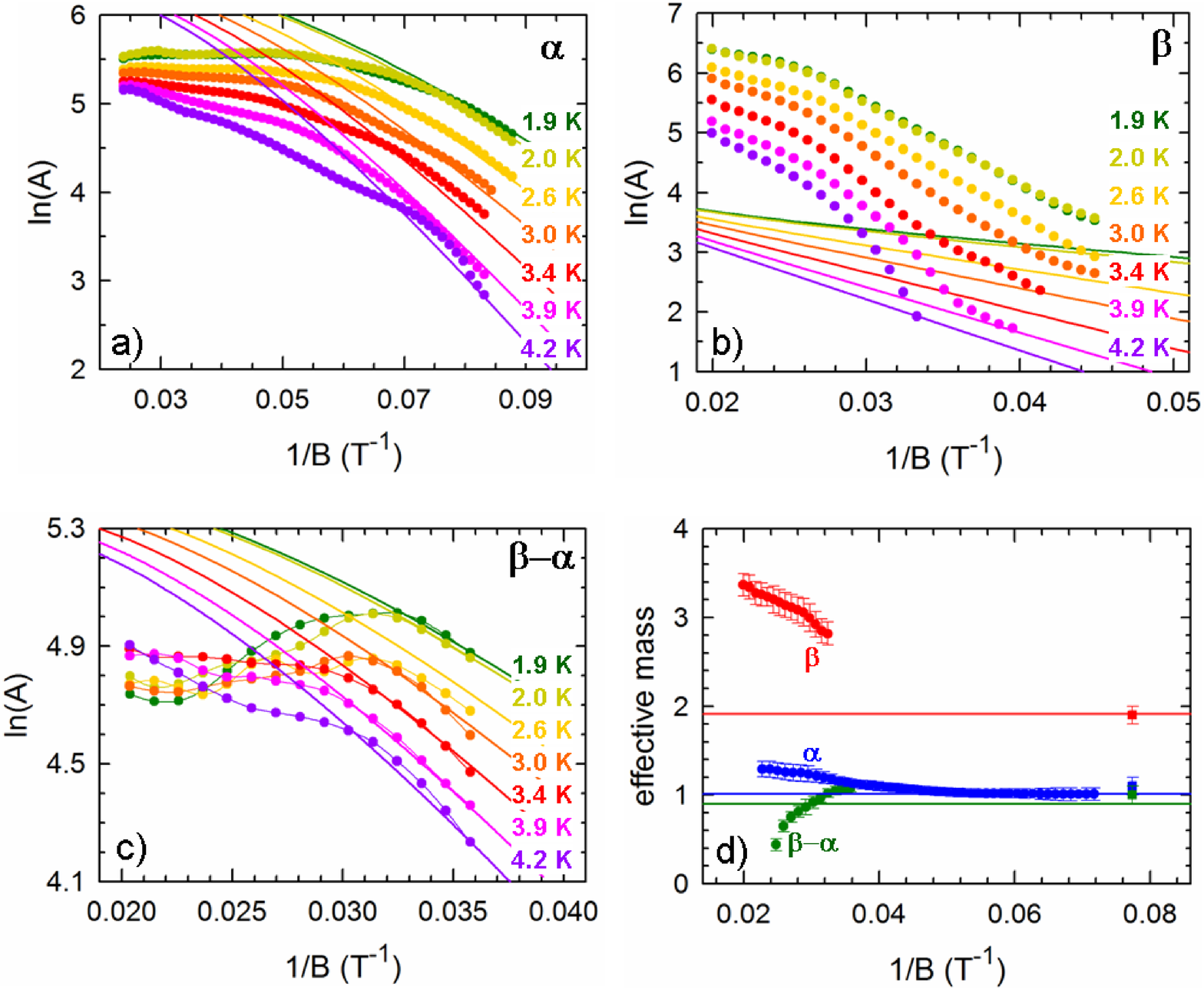}
\caption{\label{Fig:Dingle_masses_TDO} (color on line)  Dingle plots of (a) $\alpha$, (b)  $\beta$ and (c) $\beta-\alpha$  components relevant to TDO data. Solid lines are best fits of the Lifshitz-Kosevich model to the data obtained with $m_{\alpha}(\theta=0)$ = 1.00, $m_{\beta}(\theta=0)$ = 1.88 (which are the same as those deduced from dHvA data, see Fig.~\ref{Fig:alpha_beta}),  $m_{\beta-\alpha}(\theta=0)$ = 1.03 and $T_D$ = 2 K. Thin lines in (c) are guides to the eye. (d) Field dependence of the effective masses deduced from mass plots (not shown). Solid circles are deduced from the data in (a)-(c), solid squares are the values reported in Ref.~\onlinecite{Ly13}. Horizontal lines correspond to the effective mass values $m_{\alpha}$, $m_{\beta}$ and $m_{\beta}-m_{\alpha}$ deduced from dHvA data of Fig.~\ref{Fig:alpha_beta}. Solid lines are obtained with the same effective masses and Dingle temperature as in panels (a)-(c). }
\end{figure}

Once effective masses, Dingle temperature and MB field are determined, frequency combinations can be considered. As evidenced in the few cases reported as examples in Fig.~\ref{Fig:combinaisons}, data are nicely accounted for by the equations given in the appendix, with $g_{\alpha}$ = $g_{\beta}$ = 1.85$\pm$0.05, which is just the value obtained for ET$_4$-Zn \cite{Au15}. First, data for the 'forbidden frequency' $\beta-\alpha$, which only involve second order terms, is accounted for by the model. Next, strong deviation from the LK behaviour is noticed for the component $\beta+\alpha$ in Fig.~\ref{Fig:combinaisons}(b). This behaviour, already observed for ET$_4$-Zn, is due to field- and temperature-dependent cancelation of the first and second order terms of Eq.~\ref{Eq:beta+alpha} in which the second order term dominated by the product $R_{\alpha,1}R_{\beta,1}$ come close to the first order term, dominated by $R_{\beta+\alpha,1}$. In the present case, a minimum amplitude can be inferred at a temperature below the explored range, whereas the minimum takes place around 2.5-3 K for ET$_4$-Zn in the field range 47-50 T \cite{Au15}.

To go further, it can be noticed in Fig.~\ref{Fig:d_TF_FS} that the amplitude of $2\alpha$ is very small, hampering any data analysis in this case. In contrast, the amplitude of 2$\beta$ is even larger than that linked to the basic orbit $\alpha$. Contribution of the second order terms of Eqs.~\ref{Eq:2alpha} and~\ref{Eq:2beta} is directly responsible for these features. Regarding  $2\alpha$, its behaviour is due to the almost cancelation of the first and second order terms which are dominated by  $R_{\alpha,2}$ and $R_{\alpha,1}^2$, respectively (see Eq.~\ref{Eq:2alpha}). Putting aside the spin damping factors ($R^{s}_{\alpha,2}$ and $R^{s}_{\alpha,1}$), these two factors are close to each other (they are actually equal as $T/B$ goes to infinity). Owing to the tilt angle $\theta$=10$^{\circ}$, $R^{s}_{\alpha,2}$ = 0.90$\pm$0.03 is very close to $R^{s}_{\alpha,1}$$^2$ = 0.95$\pm$0.03. Hence, taking into account the spin damping factors, the first and second order terms, which enter Eq.~\ref{Eq:2alpha} with an opposite sign, keep close values and have the same sign which accounts for the observed very small amplitude. This feature is at variance with many two-dimensional organic metals, in particular with ET$_4$-Zn and $\kappa$-(ET)$_2$Cu(SCN)$_2$ for which $R^{s}_{\alpha,2}$ is negative due to larger effective mass. In contrast, a sizeable contribution of the second order terms of Eq.~\ref{Eq:2beta} relevant to 2$\beta$ is observed in Fig.~\ref{Fig:combinaisons}(a). This is mainly due to a much smaller value of the spin damping factor $R^{s}_{\beta,2}$ compared to $R^{s}_{\beta,1}$$^2$ ($R^{s}_{\beta,2}$/$R^{s}_{\beta,1}$$^2$ = 0.2 for $g_{\beta}$ = 1.85) entering Eq.~\ref{Eq:2beta}.

These findings lead us to discuss the absolute values of the Fourier amplitudes, in which spin damping factors, hence effective Land\'{e} factors play a key role. Since all the Fourier components are known within a constant factor ($\tau_0$), ratios $A_{\eta}/A_{\beta}$ are considered instead in the following. Fig.~\ref{Fig:Aeta_Abeta}, in which shaded areas account for the uncertainties on the effective masses, MB field and  Dingle temperature, display the Land\'{e} factor dependence (where it is assumed that $g_{\alpha}$ = $g_{\beta}$, see above) of such ratios calculated from Eqs.~\ref{Eq:alpha} to~\ref{Eq:2beta-alpha}. As it can be observed, values in agreement with experimental data are obtained for $g_{\alpha} = g_{\beta} =$ 1.85 $\pm$ 0.15, in nice agreement with the value deduced from the field and temperature dependence of the amplitudes (see fig.~\ref{Fig:combinaisons}) albeit with a larger uncertainty.


Turn on now on SdH oscillations which are observed in TDO data. The main feature of these data is the number of frequency combinations observed in Fig.~\ref{Fig:d_TF_FS}, even larger than for magnetic torque data. Dingle plots for $\alpha$ are displayed in Fig.~\ref{Fig:Dingle_masses_TDO}(a). Solid lines in this figure are best fits of the LK formula to the data in the low field range (keeping in mind that, as reported above, the LK model holds for the $\alpha$ and $\beta$ components amplitude of dHvA spectra). They are obtained with the effective masses and MB field derived from the dHvA oscillations and $T_D$ = 2K (remember that $T_D$ is the only sample-dependent parameter). Even though the field dependence is accounted for by the model in the low field range, strong deviations are noticed as the magnetic field increases. This behaviour, which is even more pronounced for $\beta$ (see Fig.~\ref{Fig:Dingle_masses_TDO}(b)) results in apparent field-dependent effective masses displayed in Fig.~\ref{Fig:Dingle_masses_TDO}(d), which tend towards the values derived from both the above dHvA and the low field magnetoresistance data of Ref.~\onlinecite{Ly13} as the magnetic field decreases. Noticeably, the low field part of the TDO data relevant to $\beta-\alpha$ is accounted for by $m_{\beta-\alpha}$ = 1.0 $\pm$ 0.2 which is close to $m_{\beta}-m_{\alpha}$ = 0.88 $\pm$ 0.13, hence compatible with QI. This feature confirms once again \cite{Au12,Au15} that the TDO technique is actually sensitive to conductivity rather than magnetization. This being said, not to mention QI oscillations, the analytical model which account for dHvA amplitudes is clearly not suitable for SdH oscillations at high field which still require a specific model.

\section{Summary and conclusion}

As expected for compounds with FS illustrating the linear chain of coupled orbits, dHvA and SdH spectra of  $\theta$-(BETS)$_4$CoBr$_4$(C$_6$H$_4$Cl$_2$) are composed of many linear combinations of the frequencies linked to the $\alpha$ and $\beta$ orbits. Compared to previously studied $\theta$-phase organic metals ET$_4$-Co \cite{Au12} and ET$_4$-Zn \cite{Au15}, smaller effective masses ($m_{\alpha}$ = 1.00$\pm$0.05, $m_{\beta}$ = 1.88$\pm$0.08) and MB field ($B_0$ = 11.6$\pm$3.2 T) are observed, allowing the observation of many frequency combinations in dHvA and SdH spectra in high  magnetic fields.

As already reported for other compounds with the same FS topology, analytical formulas reported in the appendix, which are based on a second order development of the free energy within the canonical ensemble, account for the field and temperature dependence of the dHvA amplitudes with Land\'{e} factors equal, within error bars, to that derived from dHvA data of ET$_4$-Zn ($g_{\alpha}=g_{\beta}$=1.85$\pm$0.05). In particular, besides the 'forbidden frequency' $\beta-\alpha$ amplitude, the non-monotonic behaviour of $\beta+\alpha$ is nicely reproduced.

Beyond the field and temperature dependence of the amplitude, the strong influence of the spin damping factor, hence of the  Land\'{e} factors, on the absolute value of the amplitudes is emphasized. In that respect, specific behaviours due to small effective masses such as the large amplitude of 2$\beta$ and the small amplitude of $2\alpha$  compared to that linked to the basic orbit $\beta$ are quantitatively interpreted.

In contrast, the analytical model suitable for dHvA amplitudes cannot account for magnetoresistance oscillations measured by TDO technique at high field. A specific model is therefore still required for SdH oscillations of the linear chain of coupled orbits.

\acknowledgements
Work in Toulouse was supported by the European Magnetic Field Laboratory (EMFL). Support of the project of Presidium RAS 0089-2015-0144 is acknowledged.

\appendix*
\section{Analytical expressions of Fourier amplitudes}
\label{analytical}

Analytical expressions of dHvA amplitudes relevant to the linear chain of coupled orbits \cite{Au12,Au15,Au16} are recalled in this appendix. Fourier amplitude $A_{p\eta}$ of the component with frequency  $F_{p\eta}=p(n_{\beta}F_{\beta}\pm n_{\alpha}F_{\alpha} )$, where $n_{\alpha(\beta)}$ is the number of $\alpha(\beta)$ orbits involved in the orbit $\eta$ and $p$ is the harmonic number, depends on  expressions involving damping factors $R_{\eta,p}(B,T)$ = $R^T_{\eta,p}(B,T) R^{D}_{\eta,p}(B) R^{MB}_{\eta,p}(B) R^{s}_{\eta,p}$, given by the LK and coupled orbits network models \cite{Fa66,Sh84}. The temperature damping factor is expressed as $R^{T}_{\eta,p} = pu_{\eta} \sinh^{-1}(pX_{\eta})$, where  $u_{\eta}$ = $u_0 m_{\eta} T/B\cos\theta$,  $u_0$ = 2$\pi^2 k_B m_e(e\hbar)^{-1}$ = 14.694 T/K, $\theta$ is the angle between the magnetic field direction and the normal to the conducting plane and $m_{\eta}=n_{\alpha}m_{\alpha}+n_{\beta}m_{\beta}$ is the effective mass. The Dingle factor is given by $R^{D}_{\eta,p} = \exp(-pu_0m_{\eta}T_D/B\cos\theta)$, where  $T_{D}$ = $\hbar(2\pi k_B\tau)^{-1}$ is the Dingle temperature and $\tau^{-1}$ is the scattering rate. MB contribution is accounted for by $R^{MB}_{\eta,p} = (ip_0)^{n^t_{\eta}}(q_0)^{n^r_{\eta}}$ where  $B_0$ is the MB field in which the tunneling ($p_0$) and reflection ($q_0$) probabilities are given by $p_0$ = $\exp(-B_0/2B\cos\theta)$ and $p_0^2$ + $q_0^2$ = 1. Finally, the spin damping factor is given by $R^{s}_{\eta,p} = \cos(\pi g_{\eta}m_{\eta}/2\cos\theta)$, where $g_{\eta}$ is the effective Land\'{e} factor.

\begin{eqnarray}
\label{Eq:alpha}
A_{\alpha}&=&\frac{F_{\alpha}}{\pi m_{\alpha}}R_{\alpha,1}+
\frac{F_{\alpha}}{2\pi m_{\beta}} R_{\alpha,1}R_{\alpha,2} + \cdots
\\ 
\label{Eq:beta}
A_{\beta}&=&\frac{F_{\beta}}{\pi m_{\beta}}R_{\beta,1}+
\frac{F_{\beta}}{\pi m_{\beta}} R_{\alpha,1}R_{\beta+\alpha,1} + \cdots
\\ 
\label{Eq:2alpha}
A_{2\alpha}&=&-\frac{F_{\alpha}}{2\pi m_{\alpha}}R_{\alpha,2}+
\frac{F_{\alpha}}{\pi m_{\beta}}\left [
R_{\alpha,1}^2-\frac{2}{3}R_{\alpha,1}R_{\alpha,3}\right ] + \cdots
\\ 
\label{Eq:2beta}
A_{2\beta}&=&-\frac{F_{\beta}}{2\pi m_{\beta}}\left [R_{\beta,2}+2R_{2\beta,1}\right ]+
\frac{F_{\beta}}{\pi m_{\beta}} \left [R_{\beta,1}^2+2R_{\alpha,1}R_{2\beta-\alpha,1}\right ]+\cdots
\\ 
\label{Eq:beta-alpha}
A_{\beta-\alpha}&=&-\frac{F_{\beta}-F_{\alpha}}{\pi m_{\beta}}\left [
R_{\alpha,1}R_{\beta,1}+\ff R_{\alpha,2}R_{\alpha+\beta,1} \right ]+ \cdots
\\ 
\label{Eq:2beta-2alpha}
A_{2(\beta-\alpha)}&=&-\frac{2(F_{\beta}-F_{\alpha})}{\pi m_{\beta}}
\left [2R_{\alpha,1}R_{2\beta-\alpha,1}+R_{\alpha,2}(R_{2\beta,1}+\ff R_{\beta,2})\right ]+ \cdots
\\ 
\label{Eq:beta+alpha}
A_{\beta+\alpha}&=&-\frac{F_{\beta}+F_{\alpha}}{\pi (m_{\beta}+m_{\alpha})}
R_{\beta+\alpha,1}+\frac{F_{\beta}+F_{\alpha}}{\pi
m_{\beta}}R_{\alpha,1}(R_{\beta,1}-R_{\beta+2\alpha,1})+ \cdots
\\
\label{Eq:2beta-alpha}
A_{2\beta-\alpha}&=&-\frac{2F_{\beta}-F_{\alpha}}{\pi (2m_{\beta}-m_{\alpha})}
R_{2\beta-\alpha,1}-
\frac{2F_{\beta}-F_{\alpha}}{\pi m_{\beta}} R_{\alpha,1}(\ff R_{\beta,2}+R_{2\beta,1})+ \cdots
\end{eqnarray}

It can be noticed that the terms of first order in damping factors correspond to the LK model. The minus signs account for $\pi$ dephasing at turning points \cite{Au14}.  With regards to Eq.~\ref{Eq:2beta}, while $R_{\beta,2}$ stands for the second harmonic of $\beta$, $R_{2\beta,1}$ is the damping factor of a MB orbit with frequency $F_{2\beta}$ as discussed in Ref.~\onlinecite{Au14}. The same spin damping factor holds for both of them.

Second order terms relevant to the Fourier component $F_{n_{\beta}\beta \pm n_{\alpha} \alpha}$ arise from an infinite series of damping factors product $R_{\eta_ 1,p_1}R_{\eta_2,p_2}$ where $|p_1\eta_1 \pm p_2\eta_2| = n_{\beta}\beta \pm n_{\alpha} \alpha$. In Eqs.~\ref{Eq:alpha} to~\ref{Eq:2beta-alpha}, only the very first terms with largest damping factors, which are not insignificant are reported.


 \end{document}